\documentclass[prd,twocolumn,amsmath,amssymb,nofootinbib,preprintnumbers,balancelastpage]{revtex4}
\pdfoutput=1
\usepackage{graphicx}
\usepackage{hyperref}
\hypersetup{
    pdfnewwindow=true,      
    colorlinks=true,       
    linkcolor=black,          
    citecolor=blue,        
    filecolor=blue,      
    urlcolor=blue           
}
\begin{document}
\title{\Large {\bf{Baryon Asymmetry and Dark Matter Through the Vector-Like Portal}}}
\author{Pavel Fileviez P\'erez}
\email{fileviez@caltech.edu}
\affiliation{Particle and Astro-Particle Physics Division \\
Max-Planck Institute for Nuclear Physics (MPIK) \\
Saupfercheckweg 1, 69117 Heidelberg, Germany}
\author{Mark B. Wise}
\email{wise@theory.caltech.edu}
\affiliation{California Institute of Technology, Pasadena, CA 91125, USA}
\date{\today}
\begin{abstract}
A possible connection between the cosmological baryon asymmetry, dark matter and vector-like fermions is investigated. In this scenario an asymmetry generated 
through  baryogenesis or leptogenesis (in the vector-like matter sector) connects the baryon asymmetry to the dark matter density. We present explicit renormalizable 
models where this connection occurs. These models have asymmetric dark matter and a significant invisible Higgs decay width to dark matter particles is possible. 
We refer to this type of scenario as the vector-like portal. In some asymmetric dark matter models there are potential naturalness issues for the low energy effective theory. 
We address that issue in the models we consider by starting with a Lagrangian that is the most general renormalizable one consistent with the gauge (and discrete) 
symmetries and showing the low energy effective theory automatically has the required form as a consequence of the symmetries of the full theory. 
We show that the mass of the dark matter candidate is predicted in these scenarios.   
\end{abstract}
\maketitle
\section{Introduction}
Two striking aspects of our universe are the baryon asymmetry 
and the dark matter relic density. There are several appealing mechanisms to explain 
 the baryon asymmetry and different dark matter candidates. However, the fact that the 
contribution to the cosmological energy density from  dark matter is not far from that of the baryons 
suggests a connection between them. 

Models that exhibit a connection between the baryon asymmetry and dark matter density 
typically (but not always) have a dark matter particle with mass around $5$ GeV and go by the name asymmetric dark 
matter since the dark matter relic density 
is determinated by an asymmetry in the dark matter anti-dark matter number densities. 
Models of this type often employ non-renormalizable couplings which leaves part of the dynamics 
relevant for explaining the cosmological density unexplored. For several scenarios discussed in the literature 
see Refs.~\cite{AD1,AD2,AD3,AD4,AD5,AD6,AD7,AD8,AD9,AD10,AD11,AD12,AD13,AD14,AD15,AD16,AD17,AD18,AD19}.
For models with heavy asymmetric dark matter see Refs.~\cite{Falkowski:2011xh, Chun:2011cc,Arina}.

In this paper we investigate simple models with vector-like fermions where an asymmetry generated 
in the vector-like sector through baryogenesis or leptogenesis is transmitted to the standard model baryons  and the 
dark matter using renormalizable coupling of the vector-like matter to the standard model fermions and the dark matter. We present a few explicit models where this connection occurs. For a range of couplings and masses these models are consistent with particle physics and cosmological constraints. They have asymmetric dark matter  and a significant 
 invisible Higgs decay width to dark matter particles is possible. We refer 
to this type of scenario as the vector-like portal. 

\begin{figure}[t] 
\includegraphics[scale=0.3]{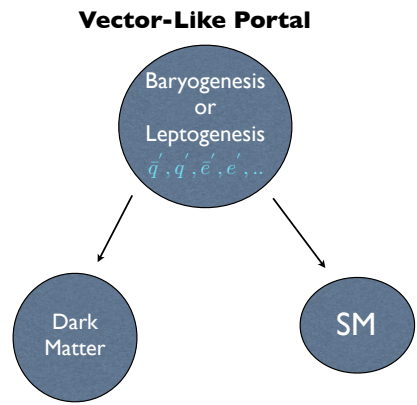}
\caption{This graph illustrates the main idea proposed in this paper, the generation of dark matter and baryon asymmetries through the vector-like portal. 
The new fermions, $q^{\prime}$, $\bar{q}^{\prime}$, $e^{\prime}$, $\bar{e}^{\prime}$, represent the vector-like fermions needed to realize this mechanism.}
\label{VLPortal}
\end{figure}

In the models presented in this article the dark matter asymmetry is determined  by an asymmetry in the charge of a new symmetry of the low energy effective theory  while the final 
baryon asymmetry is determined by the primordial $B-L$ asymmetry.  The main idea of this article is shown in Fig.~1 and we will discuss in detail the full mechanism including the  generation of the primordial asymmetries and provide an understanding of why the symmetry associated with the dark matter density exists in the effective low energy effective theory given that it must be broken to generate its primordial asymmetry. For example, if the dark matter field is a complex scalar we will need to understand why in the low energy effective theory $X^{\dagger} X$ is allowed but $X X$ is not even though in the full theory the symmetry $X \rightarrow e^{i{\alpha}}X$ must be broken to generate the primordial $X$ asymmetry. This potential naturalness problem is particularly acute for scalar dark matter\footnote{There are ways to address this issue that are different from our approach. See, for example, ~\cite{Cui:2011qe}.} .

This article is organized as follows: In section II we discuss the low-energy sectors of renormalizable models where one can achieve 
the connection between the baryon asymmetry and the dark matter density using the vector-like portal. In section III we discuss the connection between primordial asymmetries and the dark matter and baryonic asymmetries. Direct detection and Higgs decays are discussed in section IV. We show in section V how the these models can arise from the spontaneous breaking of local symmetries. 
In section VI we show how the required primordial asymmetries at very high temperature are generated. We briefly summarize our main results in section VII.
%
\section{Theoretical Framework}
In this paper we explicitly present a way to generate the baryon and dark matter asymmetries through the vector-like portal. 
We  construct  theories based on a local gauge symmetry broken at a very high scale, where primordial asymmetries are generated through the decays of  very heavy fields. The asymmetry generated in the vector-like fermionic sector is transmitted to the dark matter and the standard model through renormalizable interactions.  One of the goals of our work is to show that it is possible understand how the symmetries of the low energy effective theory  arise naturally even though they cannot be exact. 

In this section we present the low energy sectors of these models  where new vector like fermions play a role relating 
 the baryon and the dark matter densities. For simplicity we focus on cases where:
\begin{itemize}
\item There are additional vector-like quarks (leptons) that couple at tree level to the ordinary 
quarks (leptons) and the dark matter candidate. The dark matter has spin zero since then there are renormalizable coupling of this type.
\item A dark matter annihilation channel with a large rate exists so that the cosmological thermal density of 
dark matter would be negligible (compared with the observed dark matter density) if there was 
no asymmetry in the dark matter - anti dark matter number density.
\item Neutrinos are Dirac fermions.
\item We add only a few new fields beyond those in the standard model and the right handed neutrinos. 
They are needed to have the connection between the baryon asymmetry and the dark matter density. 
None of exotic particles associated with these new fields is stable except the dark matter.
\end{itemize}     
Following these guidelines we find the following simple models:
\\
\\
{\bf{Model 1}}: The Up-Quark Portal

In addition to the standard model particles and the three right handed neutrinos $\nu_R^i$, $i=1,2,3$, one has 
a gauge singlet complex dark matter scalar field X,  a vector like pair of quarks, $u_L^{\prime}$ and $u_R^{\prime}$, 
which connect the dark matter and the standard model quarks. In order to have a large annihilation rate for $X$ 
we include a gauge singlet complex scalar field $S$ which decays into two right-handed neutrinos. Therefore, the relevant Lagrangian of this discussion 
contains the following terms:
\begin{eqnarray}
{\cal L}^{(1)} & \supset &  {\cal{L}^{\prime}} - \left(M_{u^{\prime}}  \overline{u_L^{\prime}} u_R^{\prime} + \lambda_u X \overline{u}_R u_L^{\prime} + \text{h.c.}\right),
\end{eqnarray}
where in addition to kinetic terms one has
\begin{eqnarray}
{\cal L}^{'} & \supset & - m_X^2 X^\dagger X  -  \lambda_X  (X^\dagger X)^2 - m_S^2 S^\dagger S \nonumber \\
& - &  \lambda_S  (S^\dagger S)^2 - \lambda_{XS}  S^\dagger S X^\dagger X - \lambda_{HX}H^{\dagger} H X^{\dagger}X \nonumber \\ 
& - & \lambda_{HS} H^{\dagger} H S^{\dagger}S - \left( Y_\nu \overline{\ell}_L \tilde{H} \nu_R  + \lambda_R  S \nu_R \nu_R + \text{h.c.}\right). \nonumber \\
\end{eqnarray}

The coupling $\lambda_{SX}$ allows for the annihilation of $X$ into two $S$ fields, which later decay into two right-handed neutrinos. 
Even though the vector like quarks are  heavy\footnote{Their masses are assumed to be greater than a few TeV.} the coupling $\lambda_{HX}$ provides the familiar Higgs portal for direct detection experiments as well as giving rise to an invisible Higgs decay width to dark matter particles. The coupling $\lambda_{HS}$ also permits invisible Higgs decay to $S$ particles.

The model has a global $B-L$ symmetry and an exotic global $U(1)$ symmetry (with a charge we denote by $\eta$) 
where only the vector like quarks and the dark matter fields transform nontrivially corresponding to the charge assignments: 
$\eta(X)=1$,  and $\eta({u_R^{\prime}})= \eta({u_L^{\prime}})=-1$. 
This new symmetry has a discrete ${\cal Z}_2$ subgroup which is  $-1$ for $X$, $u_L^{\prime}$ and $u_R^{\prime}$, and  
 all the other fields are invariant. This discrete symmetry, like $R$-parity in supersymmetric models, ensures the stability of the dark matter. The model also has another discrete symmetry where $S \rightarrow -S$ and every lepton field, ${\ell} \rightarrow i {\ell} $.

Here the neutrinos are Dirac fermions and the Yukawa couplings $Y_{\nu}$ are very small. As we will discuss later this model 
can be obtained as the low energy limit of  a model where one has a local symmetry broken at the high scale and the most general renormalizable couplings are present. 
See the next section for details.
\\
\\
{\bf{Model 2}}: The Down-Quark Portal

The interaction between the dark matter candidate with spin zero and the 
standard model down quarks can also be used to make the connection between the baryon asymmetry and the dark matter. 
Therefore, following the same idea as in model 1 we can have the following interactions:
\begin{eqnarray}
{\cal L}^{(2)} & \supset &  {\cal{L}^{\prime}} - \left(   M_{d^{\prime}}  \overline{d_L^{\prime}} d_R^{\prime} + \lambda_d X \overline{d}_R d_L^{\prime} + \text{h.c.}\right)
\end{eqnarray}
Here we also have an ${\cal Z}_2$ symmetry which keeps $X$ stable, i.e. ${\cal Z}_2: X \to -X, d_R^{\prime} \to - d_R^{\prime},  d_L^{\prime} \to - d_L^{\prime} $.
\\
\\
{\bf{Model 3}}: The Charged-Lepton Portal

One can use also the interactions between the dark matter candidate with spin zero and the standard model charged leptons. In this case the relevant Lagrangian is given by
\begin{eqnarray}
{\cal L}^{(3)} & \supset &  {\cal{L}^{\prime}} - \left( M_{e^{\prime}}  \overline{e_L^{\prime}} e_R^{\prime} + \lambda_e X \overline{e}_R e_L^{\prime} + \text{h.c.}\right).
\end{eqnarray}
One could consider models where the dark matter has spin one-half, but in this case one needs to use non-renormalizable operators to have a large annihilation cross section 
or use the annihilation through a resonant. Since we are mainly interested in renormalizable models we proceed with the study of dark matter candidates with spin zero. 

The Yukawa couplings $Y_{\nu}$ are so small that they are ineffective at establishing thermal equilibrium at the time of
$S$ decay to the right handed neutrinos.  Hence the right handed neutrinos are extra light degrees of freedom that contribute to the expansion rate of the universe. 
 Ref.~\cite{Anchordoqui:2012qu} determined the  contribution of right-handed neutrinos to the effective 
number of relativistic degrees of freedom, $\Delta N_{eff}$, as a function of the decoupling temperature. As one can appreciate from Fig. 2 
of Ref.~\cite{Anchordoqui:2012qu} their impact on the expansion rate of the universe gets significantly diluted by the shrinking of the number of degrees of freedom during the QCD phase transition.  Therefore, provided the scalar masses of the dark matter $X$ and the scalar $S$ are are large enough (e.g. greater than 1 $\rm GeV$) the scenarios we discuss in this article are compatible with the constraints on the expansion rate of the universe coming from cosmology.

If the vector like quarks or leptons have masses in the TeV range they can give rise to some interesting signals at the Large Hadron Collider. 
In the case of the vector-like quarks one can have the QCD pair production of the new quarks and from their decays one 
gets two quarks and missing energy $$pp \ \to \ \bar{q}^{\prime} q^{\prime} \  \to \   \bar{q} q \  \rm{E_T^{miss}}.$$ 
See for example the studies in Ref.~\cite{Q-LHC} for recent bounds. In the case of vector-like leptons one has the production 
through the electroweak interactions and the signals with two leptons and missing energy
$$pp \ \to \ \bar{e}^{\prime} e^{\prime} \  \to \   \bar{e} e \  \rm{E_T^{miss}}.$$ Bound on vector-like lepton masses from the LHC are not very strong. Nonetheless in this article we will assume for simplicity that the vector-like fermions are heavy in order to avoid any severe constraints coming from charged lepton flavor violation. Of course there is no particular reason that the vector like quarks to be at the TeV scale. If they are very heavy the low energy effective theory then consists of the dark matter field $X$, the scalar field $S$ that the dark matter annihilates to and finally the standard model with Dirac neutrino masses (and hence the additional right handed neutrinos).
\section{Baryon and Dark Matter Asymmetries}
The  connection between the baryon asymmetry 
and the dark matter relic density in the models discussed above is investigated in this section. 
In this analysis we will assume that the sphalerons are rapid only above the electroweak phase transition. 
We will discuss the first model in detail and our final results hold in all the models for reasons that will become
apparent. 

Assuming that the chemical potential is much smaller than the temperature, $\mu << T$, 
one can write the asymmetry between particle and antiparticles as
\begin{equation}
\frac{n_+ - n_-}{s}=\frac{15 g}{2\pi^2 g_{*} N} \frac{\mu}{T},
\end{equation}
where $g$ counts the internal degrees of freedom, $s$ is the entropy density, and $g_{*}$ 
is the total number of relativistic degrees of freedom. Here $N=2$ for fermions and $N=1$ for bosons.
Using the fact that the third component of the isospin vanishes one has $\mu_W=0$~\cite{Harvey}, and 
we have the following conditions from the Yukawa couplings and gauge interactions:
\begin{eqnarray}
\mu_{u_L} &=& \mu_{d_L}, \  \  \mu_{e_L}=\mu_{\nu_L}, \  \ \mu_{+}=\mu_0, \nonumber \\
 \mu_{u_R} &=&  \mu_0 + \mu_{u_L}, \ \  \mu_{d_R}=-\mu_0 + \mu_{u_L}, 
\end{eqnarray}
Using the sphaleron condition
\begin{equation}
3 \left(  \mu_{u_L} + 2 \mu_{d_L} \right) \ + \ 3 \mu_{\nu_L}=0,
\end{equation}
and the conditions above we find $\mu_{\nu_L}=-3 \mu_{u_L}$. Notice that these results are valid for all models.

Let us investigate the first model. Using the condition that the electric charge is zero, $Q=0$, one gets the following equation
\begin{eqnarray}
& & 6 \left( \mu_{u_L} + \mu_{u_R} \right) \ - \ 3  \left( \mu_{d_L} + \mu_{d_R} \right) \nonumber \\ 
&- & 3  \left( \mu_{e_L} +  \mu_{e_R} \right) + 2 \mu_0 + 2  \left( \mu_{u_L^{\prime}} + \mu_{u_R^{\prime}} \right)=0. 
\end{eqnarray}
Therefore one can find the solution for the chemical potential for the new up quarks 
\begin{equation}
\mu_{u_L^{\prime}} = \mu_{u_R^{\prime}}= - 6 \mu_{u_L} -\frac{7}{2}  \mu_{0}.
\end{equation}
As we have discussed before, this model has two global symmetries with charges $\eta$ and $B-L$. Since the Yukawas
$Y_{\nu}$ are extremely small we prefer not to use in this analysis the $B-L$ that is an exact symmetry of the model but a $B-L$ where the right handed neutrinos and $S$ don't transform. Note that since the vector like quarks decay to ordinary quarks including their contribution correctly gives us the $B-L$ carried by standard model fields after their decay. 
We use the conservation of these currents  to study the connection between 
the dark matter number density asymmetry and the baryon number density asymmetry right after the decay of the vector-like quarks. One can write the 
conserved number densities as
\begin{eqnarray}
\Delta \eta &=& \frac{15}{4 \pi^2 g_{*} T} \left( 2 \mu_{X} \ - \  3  \mu_{u_L^{\prime}} - 3  \mu_{u_R^{\prime}} \right) \nonumber \\
&=& \frac{15}{4 \pi^2 g_{*} T} \left(  50 \ \mu_{u_L} +  30 \ \mu_{0} \right),
\end{eqnarray}
and
\begin{eqnarray}
\Delta (B-L) &=& \frac{15}{4 \pi^2 g_{*} T} \left[ 3  \left( \mu_{u_L} + \mu_{u_R}  +  \mu_{d_L} + \mu_{d_R} \right) \right.  \nonumber \\ 
& + & \left. \left( \mu_{u_L^{\prime}} + \mu_{u_R^{\prime}}  \right) - 3  \left(  \mu_{e_L} +  \mu_{e_R} + \mu_{\nu_L}\right) \right] \nonumber \\
&=& \frac{15}{4 \pi^2 g_{*} T} \left(  27 \mu_{u_L} - 4 \mu_0 \right).
\end{eqnarray}
Immediately after vector-like decay the baryon number density asymmetry is
\begin{eqnarray}
B^{(1)}&=&\frac{n_B - n_{\bar{B}}}{s} = \frac{15}{4 \pi^2 g_{*} T} \left(  3 \left(  \mu_{u_L} + \mu_{d_L} \right) \right. \nonumber \\
& + & \left.  3  \left(  \mu_{u_R} + \mu_{d_R} \right)  \ + \  \mu_{u_L^{\prime}}  +   \mu_{u_R^{\prime}} \right) \nonumber \\
&=& \frac{15}{4 \pi^2 g_{*} T} (-7 \mu_0).
\end{eqnarray}
The dark matter asymmetry in this case can be written as
\begin{eqnarray}
\Delta n_X^{(1)}&=&\frac{n_X - n_{\bar{X}}}{s}= \frac{15}{2 \pi^2 g_{*} T} \left( \mu_{X}  \ - \   \frac{3}{2} \left( \mu_{u_L^{\prime}} +  \mu_{u_R^{\prime}}  \right) \right)\nonumber \\
&=& \frac{15}{2 \pi^2 g_{*} T} \left( 15 \mu_0 \ + \ 25 \mu_{u_L} \right),
\end{eqnarray}
where the last term in the second line of the above equation 
is due to the decays of $u_L^{\prime}$ and $u_R^{\prime}$. Then, using the equations 
above one can write the dark matter asymmetry as a function 
of the conserved asymmetry, $\Delta \eta$
\begin{eqnarray}
\Delta n_X^{(1)} &=&\Delta \eta.
\end{eqnarray}
The final baryon asymmetry is defined by $B-L$ as pointed out in Ref.~\cite{Harvey}, 
\begin{equation}
\Delta B_f = \frac{28}{79} \Delta (B-L).
\end{equation}
and the final value of the dark matter number density  is given by the asymmetry $\Delta \eta$.
Hence the relation between the dark matter relic density and the baryon asymmetry is,
\begin{eqnarray}
\frac{\Omega_{DM}/M_X}{\Omega_B/M_p}&=&{79 \over 28}{\bigg | }\frac{ \ \Delta \eta }{ \ \Delta (B-L)}{\bigg |}.
\label{keyrelation}
\end{eqnarray}
It is important to emphasize that Eq.~(\ref{keyrelation}) is valid for all models studied in this paper. 
The primordial asymmetry in the new sector defines the dark matter asymmetry and the primordial 
$B-L$ defines the final baryon asymmetry. In order to predict the dark matter mass one 
needs to have the full mechanism that generates the asymmetries. This is our main task in 
Section VI. 
\section{Relic density, direct detection and Higgs decays}
The relic density of asymmetry dark matter models has been studied in Ref.~\cite{Hoernisa}. 
In our case since the dark matter X has  a large annihilation cross section into two S fields, 
the relic density is set just by the asymmetry. We  estimate the annihilation cross section 
of the dark matter candidate into two $S$ fields as
\begin{equation}
\sigma v ( X X^\dagger \to S S^\dagger)=\frac{\lambda_{XS}^2}{64 \pi M_X^2} \left( 1 -  \frac{M_S^2}{M_X^2} \right)^{1/2}.
\end{equation} 
Using $M_S=0.5$ GeV, $M_X=1$ GeV, and $\lambda_{XS} \sim 1$ a large value for
$\sigma v ( X X^\dagger \to S S^\dagger) \sim 7.3 \times 10^{-20} \rm{cm^3/s}$ results. Therefore, 
the thermal relic density is very small in this case and the dark matter density is defined by the asymmetry $\Delta \eta$.

The dark matter candidate $X$ in these models has a tree level interaction with the SM Higgs. 
This coupling determines the scattering between the dark matter candidate and the nucleon, 
relevant for dark matter experiments. There is also a contribution to the scattering 
mediated by the vector-like quarks (leptons), but since these quarks should be heavy, a few TeV or more, 
the scattering through the Higgs portal dominates. The elastic DM-nucleon cross section is given by
\begin{equation}
\sigma_{SI}=\frac{ \lambda_{HX}^2}{\pi M_h^4} \frac{M_N^4 f_N^2}{ \left(  M_X + M_N \right)^2}
\end{equation}
where $\lambda_{HX}$ is the coupling between the dark matter candidate and the SM Higgs, 
$\lambda_{HX} X^\dagger X H^\dagger H$. Here $f_N$ is defined by 
\begin{equation}
<N| \sum_{q} m_q \bar{q} q |N>=f_N M_N \bar{N} N.
\end{equation}
Here we use $f_N=0.3$ for the numerical analysis. See also Ref.~\cite{Mambrini:2011ik} for a discussion of the matrix elements.
In order to estimate the elastic cross section between the nucleon and the DM candidate, 
we use $M_N=1$ GeV, $M_X=1$ GeV, $\lambda_{HX}=0.5 \times 10^{-2}$, $M_h=125$ GeV, $f_N=1/3$ and one 
finds $\sigma_{SI} \sim 3.5 \times 10^{-43} \rm{cm}^2$. Notice that the most important limits on 
the elastic cross section for light dark matter candidates are given by the Xenon10 Collaboration~\cite{Xenon10}.  
However, when the dark matter mass is below 10 GeV the best bounds are coming from CRESST-1~\cite{CRESST} 
and they are very weak.

In the models presented in this article the dark matter candidate is light and has a tree level 
coupling to the SM Higgs. Therefore, one always will have invisible Higgs 
decays as generic consequence of having a bosonic asymmetry dark matter scenario. 
The decay width of the SM Higgs into dark matter is given by
\begin{eqnarray}
\Gamma (h \to XX)&=&\frac{\lambda_{HX}^2 v^2}{8 \pi M_h} \left( 1 - 4 \frac{M_X^2}{M_h^2} \right)^{1/2}, \\
&=& \frac{\sigma_{SI} M_h^3 v^2}{8 M_N^4 f_N^2} \left(M_X + M_N\right)^2 \left(1 -  \frac{4 M_X^2}{M_h^2} \right)^{1/2}. \nonumber
\end{eqnarray} 
Now, using $M_h=125$ GeV, $v=246$ GeV, $M_N=1$ GeV, $M_X=1$ GeV, and $f_N=1/3$ one finds
\begin{eqnarray}
\Gamma_{inv} > \Gamma (h \to XX)&= & 5.3 \times 10^{11} \  \rm{GeV}^3 \times \rm{\sigma_{SI}}.
\end{eqnarray} 
Therefore, the invisible decay of the Higgs provides 
an upper bound on the spin independent cross section for direct detection.
%
\section{Model with Local Symmetry}
We have mentioned above that the models discussed in this article can be obtained from a model where one has 
an Abelian local symmetry: $U(1)_\chi$ and writes down the most general interactions consistent with the particle content and gauge symmetries.  For concreteness we just work with model 1 although similar results can be obtained for all the models. Recall that  for model 1 the Lagrangian is
\begin{eqnarray}
{\cal L} & = & {\cal L}_{\rm SM} \ + \ {\cal L}_{Kin} -   \left( Y_\nu  \bar{\ell}_L \tilde{H} \nu_R \ + \  \lambda_R S \nu_R \nu_R \right.  \nonumber \\
& + & \left. M_{u^{\prime}} \bar{u}_L^{\prime} u_R^{\prime} \ + \  \lambda_u  X \bar{u}_R u_L^{'} \ + \  \rm{h.c.}\right) \nonumber \\
&-& V(H, S, X, \chi), 
\label{lag1}
\end{eqnarray}
where the scalar potential is given by
\begin{equation}
\label{lag2}
V(H, S, X, \chi) = \sum_{i} m_{\phi_i}^2 \phi^\dagger_i  \phi_i \ + \  \sum_{i,j} \lambda_{ij} \left( \phi^\dagger_i \phi_i \right) \left( \phi_j^\dagger \phi_j \right)
\end{equation}
with $\phi_i=H,S,X,\chi$. Here the field $\chi$ gets the VEV and breaks the local $U(1)_{\chi}$ symmetry. The charges under this symmetry of all the standard model fields (including the right handed neutrinos) is just $B-L$. Hence the charge of the field $S$ is, $n_{\chi}(S)=2$. The charges of the primed vector like up quarks is the same for left and right handed fields, $n_{\chi}(u^{\prime})=n_{\chi}(u_L^{\prime})=n_{\chi}(u_R^{\prime})$. Finally, $n_{\chi} (X) = 1/3 - n_{\chi} (u^{\prime})$. In addition to this gauge symmetry we also impose the two discrete symmetries (i) $X\rightarrow -X$, $u^{\prime}_L \rightarrow -u^{\prime}_L$, $u^{\prime}_R \rightarrow -u^{\prime}_R$ and (ii) $\ell \rightarrow i \ell$ and $S \rightarrow -S$ for any lepton field $\ell$. Clearly there are many choices of the two free charges $
n_{\chi}(u^{\prime})$ and $n_\chi (\chi)$ that result in the Lagrangian above\footnote{ We need to forbid, for example, operators like $XX$ from appearing in the Lagrangian after symmetry breaking.}.
\section{Baryogenesis and Leptogenesis Mechanisms}
To complete the model and  have a complete framework we need to  show how to  generate the primordial $B-L$ and $\eta$ asymmetries. 

\subsection{ Baryogenesis Mechanism}

Following the recent paper~\cite{Arnold} (see also~\cite{Babu}) and using the interactions in model 1 we add very heavy di-quarks which decay to the vector-like quarks. 
The relevant part of the Lagrangian is given by
\begin{eqnarray}
\label{lag3}
{\cal{L}} &\supset & \lambda_1 \Delta_2 u_L^{\prime} u_L^{\prime} \ + \  \lambda_2 \Delta_2 u_R^{\prime} u_R^{\prime} 
\ + \  \lambda_3 \Delta_1 d_R d_R \nonumber \\
& + &  \lambda_4 \Delta_2 \Delta_1 \Delta_1 \chi  +  \rm{h.c.} 
\end{eqnarray}
where one needs two copies of $\Delta_2 \sim (\bar{6},1,-4/3)$, and one of $\Delta_1 \sim (\bar{6},1,2/3)$. The field $\chi$ breaks local $U(1)_\chi$ at the very high scale and generates the needed trilinear scalar term between the scalars. Now $n_{\chi}(\Delta_1)=-2/3$ and all the  charges are determined in terms of the one free charge $n_{\chi}(u^{\prime})$,
\begin{equation}
n_{\chi}(\Delta_2) = -2 n_{\chi}(u^{\prime}), {\rm{and}}~~n_{\chi}(\chi)=2n_{\chi}(u^{\prime})+4/3. 
\end{equation}
For simplicity lets consider the case where the vector like up quarks are also very heavy. Then the low energy effective theory consists of the standard model fields (with right handed neutrinos), $S$ and $X$. Since the symmetry $U(1)_{\chi}$ is spontaneously broken at a very high scale we must argue that arbitrary insertions of the vev of $\chi$ or its complex conjugate (in any Feynman diagram that is not extremely small) does not induce the operators $O_1=X^2$, $O_2=S^2$, $O_3=X^2 S^2$, $O_4=X^2(S^{\dagger})^2$, $O_5=X^4$ and $O_6=S^4$ in the low energy effective theory that results from integrating out the very heavy fields. Note we do not have to consider the operators, $ X^3$, $S^{\dagger}X^2$,  {\it etc}, because of the discrete symmetries imposed.  Operators of higher dimension than these (e.g. $X^6$ )  are suppressed by powers the very high scale.

Fortunately we do not have to examine the Feynman diagrams in the model, but rather implement conditions on the charge assignment  $n_{\chi}(u^{\prime}$) that forbid all the bad operators from being induced. The conditions we must impose are
\begin{equation}
n_{\chi}(O_j) \ne  p\left(2n_{\chi}(u^{\prime})+4/3\right) ,p=0,\pm 1, \pm 2, \ldots,
\end{equation}
for $ j=1,...,6$. The $u^{\prime}$ charge can be chosen so that these inequalities are satisfied. A simple  choice is to take $n_{\chi}(u^{\prime})=3$. This implies that, $n_{\chi}(X)=-8/3$ and $n_{\chi}(\chi)=22/3$. 

Note there is no  $U(1)_{\chi}$ charge assignment for $u^{\prime}$ that can forbid the operator $S^*X^2$ from appearing in the low energy effective theory, since $n_{\chi}(S^*X^2)=-4/3-2 n_{\chi}(u^{\prime})$, rather it does not occur because of the discrete symmetries imposed. 

The choice of matter and the gauge and discrete symmetries we have mentioned fixes the Lagrangian to be the one we have given, but then they are accidental global symmetries. The most powerful one has the charges the same as the gauge symmetry but with $n_{\chi}(u^{\prime})$ chosen to be irrational. Using this symmetry (and the discrete symmetries) it is easy to see that that the first bad non renormalizable operator that might arise from integrating out the heavy degrees of freedom  is  the dimension six operator, $S^{*2}X^4$. However, we have not managed to find a simple Feynman diagram that generates it so even if it does exist it will be severely  suppressed by coupling constants and loop factors in addition to the mass scale suppression associated with its dimension and hence such operators can be safely neglected for the evolution of the universe after the generation of the primordial asymmetries.

In this model $\Delta_2$ has the following decays 
\begin{equation}
\ \ \ \Delta_2 \to \Delta_1^* \Delta_1^*,~~~ \Delta_2 \to \bar{u}^{\prime} \bar{u}^{\prime},
\end{equation}
and using the self-energy contributions one can compute the baryon asymmetry in the vector-like sector. 
See Fig. 2 for the relevant graphs needed to compute the baryon asymmetry in the vector-like sector. 

In Table 1 we show the different channels contributing to the baryon asymmetry, including the branching ratios and the contributions to the $B-L$ and $\eta$ asymmetries.
\begin{table}[h]
\begin{tabular}{|c|c|c|c|}
\hline
~~~~~~~~ Decay ~~~~~~~~&~~~~ Br ~~~~&~~ $(B-L)$& $\eta$\\ \hline \hline
$\Delta_2 \rightarrow \Delta_1^* \Delta_1^*$ & $  1-r$ & $ 4/3$& $0$ \\ \hline
$\Delta_2 \rightarrow \bar{u}^{\prime} \bar{u}^{\prime}$ & $ r$ & $ -2/3$& $2$ \\ \hline
$\Delta_2^{*} \rightarrow \Delta_1 \Delta_1$ & $1-\bar{r}$ & $-4/3$&$0$ \\ \hline
$\Delta_2^* \rightarrow u^{\prime} u^{\prime}$ & $\bar{r}$ & $2/3$& $-2$ \\ \hline
\end{tabular}
\caption{Branching ratios and final states for the decays of $\Delta_2$ and $\Delta_2^{*}$ which contribute to the baryon asymmetry.}
\label{table1}
\end{table}

Using Table \ref{table1} we can see that this mechanism predicts the following ratio between the asymmetries
\begin{equation}
\frac{\Delta \eta}{\Delta (B-L)}=-1.
\end{equation}
Therefore, the dark matter mass in models 1 and 2 is given by
\begin{eqnarray}
M_X &=& M_p \frac{28 \  \Omega_{DM} }{79 \  \Omega_p} \approx 1.8 \ \rm{GeV}.
\end{eqnarray}
Notice that the dark matter is light as one expects. 
\begin{figure}[h] 
\includegraphics[scale=0.6]{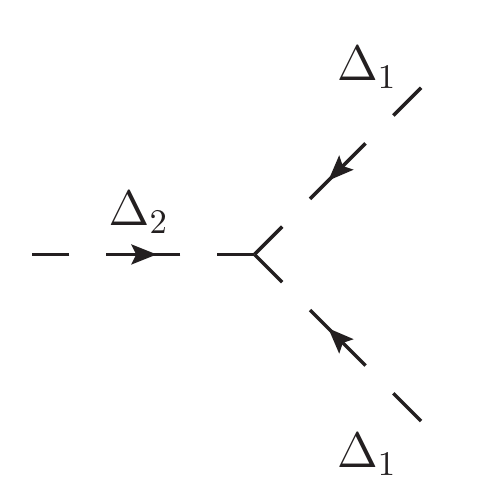}
\includegraphics[scale=0.6]{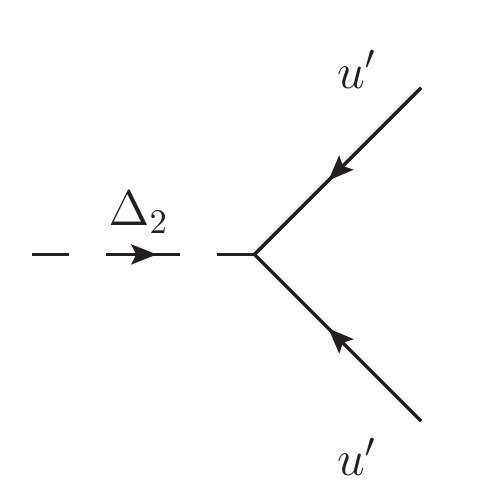}
\includegraphics[scale=0.6]{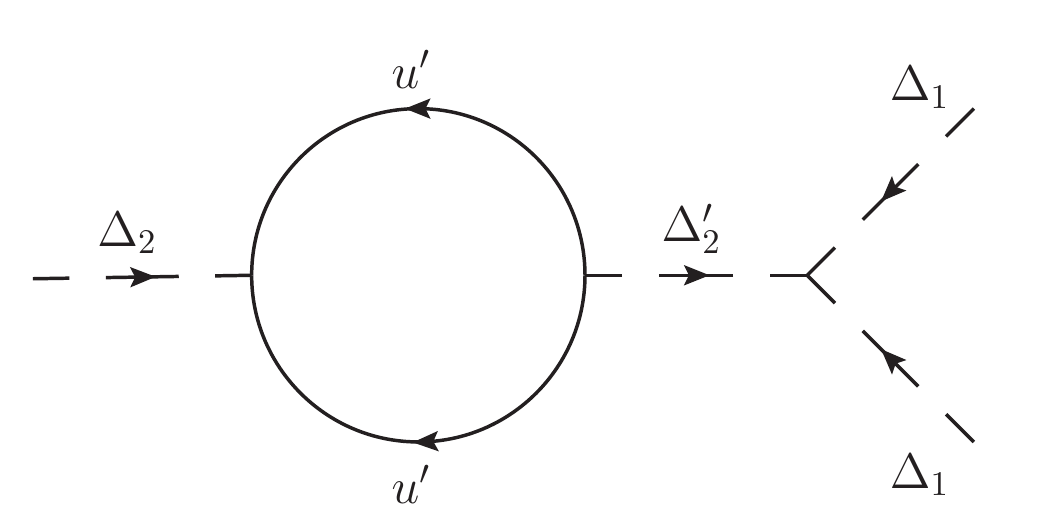}
\includegraphics[scale=0.6]{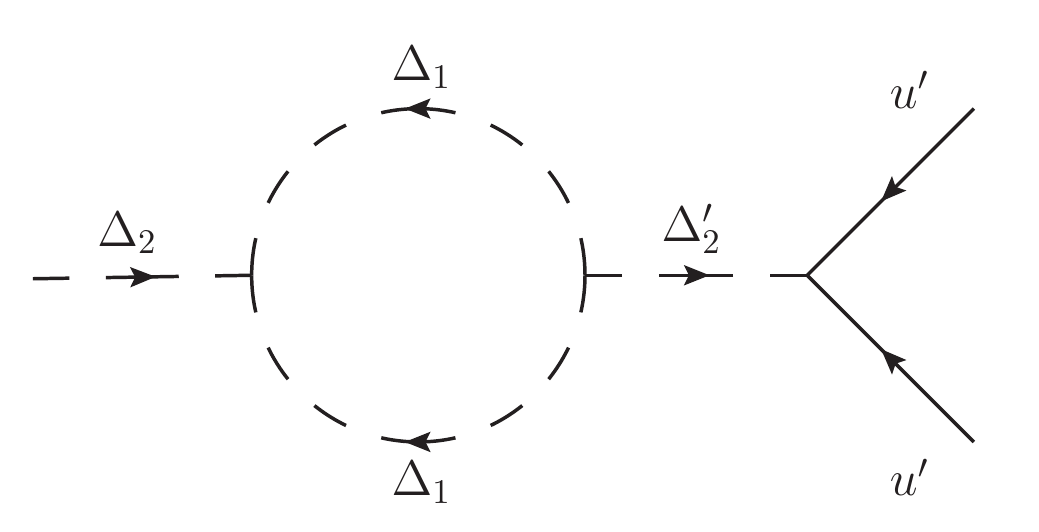}
\caption{Feynman graphs for the generation of baryon asymmetry.}
\label{graphs}
\end{figure}
This scenario is interesting because one can predict the dark matter mass.
Note we have have assumed that the decays occur out of equilibrium so that  washout effects are negligible.

We  can generate the $B-L$ asymmetry through the decays of  the heavier of the two $\Delta_2$'s (the lighter we denote
by ${\tilde \Delta}_2$ and we similarly use a tilda to denote its couplings).
 Without loss of generality we take the vacuum expectation value of $\chi$, $v_{\chi}$ to be real and the couplings $\lambda_4$ and ${\tilde \lambda_4}$  to be real. For simplicity we assume that the coupling $\lambda_2$ is much greater than the other $\lambda$'s so the dominant decay
of $\Delta_2$ is to ${\bar u_R}' {\bar u_R'}$. Furthermore we assume that the mass $M_{\Delta_2}$ is much greater than the masses of the other particles. With these simplifying assumptions and using the results from Table \ref{table1},
\begin{equation}
{\Delta (B-L) \over s} \simeq -{6 \over \pi}{ d \over g_{\star}}{v_\chi^2 \over M_{\Delta_2}^2} {\rm Im} \left[{ \lambda_4 {\tilde \lambda_4} \lambda^{*}_2 {\tilde \lambda}_2 \over | \lambda_2|^2} \right].
\end{equation}
Here $g_{\star} \sim 200$ is the number of active degrees of freedom at the time $\Delta_2$ decays and $d$ is a dilution factor that we assume is close to unity. Clearly this can be around the needed value ($\sim 10^{-10}$) even for quite small couplings.

\subsection{ Leptogenesis Mechanism}

This case is similar to the previous one and so we will be brief. In model 3 where we have vector-like leptons one can generate the asymmetry through leptogenesis 
using the following interactions,
\begin{eqnarray}
{\cal{L}} &\supset & h_1 \Delta_2  e_L^{\prime} e_L^{\prime} \ + \  h_2 \Delta_2  e_R^{\prime} e_R^{\prime} \  \nonumber \\
& + &  h_3  \Delta_1  \ell_L \ell_L \ + \ h_4 \Delta_2 \Delta_1^* \Delta_1^* \chi  +  \rm{h.c.} 
\label{leptodelta}
\end{eqnarray}
where here one needs two copies of $\Delta_2 \sim (1,1,2)$ and one of $\Delta_1 \sim (1,1,1)$. 
As in the model for baryogenesis the field $\chi$ breaks the local symmetry and one obtains 
the trilinear term between the scalars needed for leptogenesis. In this case the charges for the fields 
are $n_\chi(\Delta_1)=2$, $n_\chi(\Delta_2)=-2 n_\chi(e^{\prime})$,  where $n_\chi(e^{\prime})=n_\chi(e_L^{\prime})=n_{\chi}(e_R^{\prime})$. Furthermore  one has the conditions
\begin{eqnarray}
n_\chi (X)&=& -1- n_{\chi}(e^{\prime}), \\
n_\chi (\chi)&=& 4 + 2 n_{\chi} (e^{\prime}).
\end{eqnarray}
Following the analysis above for the baryogenesis mechanism we can avoid all bad operators 
using the condition.%
\begin{equation}
n_{\chi}(O_j) \ne  p\left(2n_{\chi}(e^{\prime})+4\right) ,p=0,\pm 1, \pm 2, \ldots,
\end{equation}
for $j=1,...,6$. 
together with some discrete symmetries. One of the discrete symmetries $X \rightarrow -X$, $e'_{R,L} \rightarrow -e'_{R,L} $. The second is an approximate discrete symmetry. Since the right handed neutrinos are Dirac their Yukawa couplings, $Y_{\nu}$  are very small they can be neglected. In this limit we impose the discrete symmetry $ S \rightarrow -S$ and $\nu_R \rightarrow i\nu_R$. It is this latter discrete symmetry that prevented the coupling  $\Delta_1 e_R \nu_R \nonumber$ from appearing in eq.~(\ref{leptodelta}).
We can generate the lepton asymmetry through the decays of  the heavier of the two $\Delta_2$'s 
\begin{equation}
\  \  \  \Delta_2 \to \Delta_1 \Delta_1, \Delta_2 \to \bar{e}^{\prime} \bar{e}^{\prime}.
\end{equation} 
One can estimate the lepton asymmetry in a similar way as discussed in the baryogenesis mechanism. 
Table II shows the branching ratios of the different channels relevant for our discussion.

\begin{table}[h]
\begin{tabular}{|c|c|c|c|}
\hline
~~~~~~~~ Decay ~~~~~~~~&~~~~ Br ~~~~&~~ $(B-L)$& $\eta$\\ \hline \hline
$\Delta_2 \rightarrow \Delta_1 \Delta_1 \rightarrow \bar{e}_L \bar{\nu}_L \bar{e}_L \bar{\nu}_L $ & $ r $ & $4$& 0 \\ \hline
$\Delta_2 \rightarrow \bar{e}^{\prime} \bar{e}^{\prime}$ & $ 1-r$ & $ 2$& 2\\ \hline
$\Delta_2^{*} \rightarrow \Delta_1^{*} \Delta_1^{*} \rightarrow e_L \nu_L e_L \nu_L$ & $ \bar{r}$ & $-4$&0\\ \hline
$\Delta_2^* \rightarrow e^{\prime} e^{\prime}$ & $1-\bar{r}$ & $-2$& -2 \\  \hline
\end{tabular}
\caption{Branching ratios and final states for the decays of $\Delta_2$ and $\Delta_2^{*}$ which contribute to the baryon asymmetry.}
\label{table2}
\end{table}
Therefore, the dark matter mass in models 3 is also given by
\begin{eqnarray}
M_X^{(3)} &=& M_p \frac{28 \  \Omega_{DM} }{79 \  \Omega_p} =  1.8 \ \rm{GeV}.
\end{eqnarray}
A detailed numerical calculation of the primordial CP-violating asymmetries $\Delta \eta$ and $B-L$ requires solving Boltzmann equations and is beyond the scope of this article. The main aim of this section was to explicitly show how one can generate the required primordial asymmetries. 
\section{Summary}
We have explored models where vector like fermions connect the dark matter and standard model sectors giving rise to a connection between
the baryon asymmetry and the dark matter density. We refer to this type of scenario as the Vector-Like Portal. In these scenarios the asymmetry generated 
in the vector-like sector through baryogenesis or leptogenesis is transmitted to the visible sector and the 
dark matter sector. The  dark matter asymmetry is always determined by the conserved asymmetry in the new sector and the baryon asymmetry is determined by the primordial $B-L$ asymmetry.

We constructed renormalizable models that realize the vector-like portal mechanism and hence all the dynamics
was explicitly displayed. Furthermore the Lagrangians we used were the most general ones consistent with the gauge (and discrete) symmetries of the theory. The complete dynamics associated with generating these cosmological densities was presented. We pointed out that the low energy effective theories obtained are natural including arbitrary loop diagrams in the full theory. This is non trivial because the symmetry of the low energy effective theory that is responsible for the conserved dark matter charge is not exact but must be broken to generate the primordial dark matter density. 

In this paper we focused on a particular way of generating  the primordial 
asymmetries using new fields that couple to the vector like fermions and ordinary fermions. This may not be the most compact or elegant way to generate the primordial asymmetries.

The predictions for direct detection have been discussed, and we found a simple connection between the invisible decay width of the SM Higgs and the spin-independent elastic 
nucleon-dark matter cross section. Since in this scenario the dark matter is always light, and the bounds coming from direct detection in this region are weak the invisible decay width of the Higgs could be large.
Finally, it is important to mention that one can predict the dark matter mass in the scenarios discussed in this article.

{\textit{Acknowledgment}}:
{\small{We thank J. Arnold and C. Cheung for discussions.
P.F.P. thanks CALTECH for the great hospitality. The work of M.B.W. was supported in part by the U.S. Department of Energy under contract No. DE-FG02-
92ER40701 and by the Gordon and Betty Moore Foundation.}

\end{document}